\documentclass[final,5p,times,twocolumn,authoryear]{elsarticle}
\usepackage[utf8]{inputenc}
\usepackage{amssymb}
\usepackage{array}
\usepackage{tabularx}
\usepackage{lipsum}
\usepackage{comment}
\usepackage{amsmath}
\usepackage[acronym,nopostdot,nonumberlist]{glossaries-extra}
\usepackage{glossary-inline}
\usepackage[table]{xcolor}
\usepackage{longtable}
\usepackage{array}
\usepackage{multirow}
\usepackage{booktabs}
\usepackage{microtype}
\usepackage{textcomp}

\usepackage{xcolor}

\renewcommand*{\glossarysection}[2][]{}

\setabbreviationstyle[acronym]{long-short}
\makeglossaries

\newacronym{uca}{UCA}{unsafe control action}
\newacronym{stpa}{STPA}{system-theoretic process analysis}
\newacronym{asil}{ASIL}{automotive safety integrity level}
\newacronym{stamp}{STAMP}{system-theoretic accident model and processes}

\newacronym{mcs}{MCS}{monte carlo simulation}

\newcommand{\abbreviations}[1]{%
  \nonumnote{\textit{Abbreviations:\enspace}#1}}

\begin{document}

\begin{frontmatter}

\title{A Scalable Framework for the Management of STPA Requirements: a Case Study on eVTOL Operations}

\author[first]{Shufeng Chen}
\affiliation[first]{organization={WMG, University of Warwick},
            addressline={6 Lord Bhattacharyya Way}, 
            city={Coventry},
            postcode={CV4 7AL},
            country={United Kingdom}}
            
\affiliation[inst2]{organization={ Graduate School of Frontier Sciences, The University of Tokyo},
            addressline={ 5-1-5 Kashiwanoha }, 
            city={Kashiwa},
            postcode={277-8563}, 
            country={Japan}}
            
\author[first]{Halima El Badaoui}
\author[first]{Mariat James Elizebeth}
\author[inst2]{Takuya Nakashima}
\author[first]{Siddartha Khastgir}

\author[first]{Paul Jennings}

\abbreviations{\printglossary[type=\acronymtype,style=inline]{}.}

\begin{abstract}
System-Theoretic Process Analysis (STPA) is a recommended method for analysing complex systems, capable of identifying thousands of safety requirements often missed by traditional techniques such as Failure Mode and Effects Analysis (FMEA) and Fault Tree Analysis (FTA). However, the absence of a structured framework for managing and prioritising these requirements presents challenges, particularly in fast-paced development environments. This paper introduces a scalable framework for prioritising STPA-derived requirements. The framework integrates outputs from each STPA step and incorporates expert evaluations based on four key factors: implementation time, cost, requirement type, and regulatory coverage. To reduce subjectivity, Monte-Carlo Simulation (MCS) is employed to calculate and stabilise requirement rankings. An automation toolchain supports the framework, enabling dynamic mapping of prioritised requirements in a scaling matrix. This visualisation aids decision-making and ensures traceability across development phases. The framework is applicable from early conceptualisation to more advanced stages, enhancing its utility in iterative system development. The framework was validated through a real-world case study focused on Electric Vertical Take-off and Landing (eVTOL) operations, conducted in collaboration with the UK Civil Aviation Authority. The findings contributed directly to CAP3141, a Civil Aviation Publication that identifies systemic operational risks and safety mitigations for regulators, operators, and vertiports. The prioritisation process supported decision-making by helping stakeholders identify and manage high-impact requirements efficiently. This work contributes a practical solution for managing STPA outputs, bridging gaps in requirement prioritisation and supporting safety-critical development in emerging technologies.
\end{abstract}


\end{frontmatter}

\section{Introduction}
\label{introduction}
With the increasing complexity of both technical and socio-technical systems, many new safety analysis methods have been developed to ensure that the safety-critical requirements due to the emergent properties of the systems are not overlooked. System-Theoretic Process Analysis (STPA) is a system-thinking-based method that originated from MIT. It is part of the System-Theoretic Accident Model and Processes (STAMP) and has been used in various studies across different \cite{sadeghi2023proposed}. Including aviation ~\cite{allison2017systems}, \cite{schmid2018laser}, medical sector~\cite{wong2020stamping,yamaguchi2019system,chen2021analyzing}, maritime~\cite{yang2020systems,ventikos2020systems,nakashima2025addressing}, automotive~\cite{zhou2019hazard,khastgir2021systems,khastgir2017introducing,chen2020identifying}, railway safety management~\cite{tonk2024application,riccardi2025structural,chen2025hierarchical} , and the safety analysis in the area of Large Language Models (LLM)~\cite{qi2025safety}. Especially in the aviation domain, STPA was highly recognized as a recommended methodology in the report \cite{thomas2024evaluation}.  It has been recognised as a recommended method to identify requirements that may not be identified by other traditional methods like Failure Mode and Effects Analysis (FMEA) and Fault Tree Analysis (FTA) \cite{comparison}.

STPA can be applied at the very early stage of the system development when available information about the system is very limited. As a deductive method, it starts with the stakeholders' engagement to identify any unacceptable losses that must be prevented (i.e., step 1). Control Structure(s) that represent the graphical interactions of the system stakeholders, in the form of nested control loops, are then created (i.e., step 2). Based on the Control Structure(s), each CA is then analysed to understand how they might manifest into Unsafe Control Actions (UCAs) (i.e., step 3). Each UCA is then further analysed, together with the analyses of the relevant control loops, to identify their possible Causal Factors (CF) and requirements to either prevent or detect these CFs (i.e., step 4).

For a complex system like the autonomous vehicle ecosystem or helicopter operation, STPA identifies up to a few thousand requirements. Although STPA has been recognized as a recommended method for complex system safety analysis, the lack of frameworks to manage and prioritise these requirements presents unique challenges, especially when applying STPA in a fast-paced development cycle. 

Requirement prioritisation is the process of determining the relative importance or urgency of requirements to guide decision-making during the development of the system. It helps stakeholders identify what to implement first, ensuring that the most valuable features are delivered early and that resources are allocated effectively. There have been many methods for requirement prioritisation developed to optimise the requirement engineering process. These include MoSCoW \cite{moscow}, (Analytic Hierarchy Process ) AHP \cite{ahp}, Cumulative Voting \cite{cumulativevoting}, and Kano Model \cite{kano}, etc. However, the limitations, such as subjectivity-oriented, lack of scalability, communication of results, and change management, still exist in these methods, making them less applicable when managing the requirements of a complex system \cite{Kivanc2025SMSRiskAviation}. Having identified these limitations of the existing requirement prioritisation techniques, in this paper, the authors aim to address the following research questions:

\subsection{Research Gaps}
\begin{itemize}
    \item RQ1: How can the requirements identified from the STPA be effectively implemented for safe system operation?
    
     
    \item RQ2: How can we objectively prioritise all the requirements identified by STPA?
    \item RQ3: How can the prioritised requirements be better communicated with the stakeholders?
\end{itemize}
\subsection{Paper Contribution} 

In this paper, a new framework to optimise the prioritisation of STPA requirements is presented. The framework was designed in a way that parameterises and utilizes the outputs from the previous steps of STPA. \textcolor{black}{Our previous work regarding the framework for prioritisation of UCAs was introduced \cite{UCAPrio}. For simplicity, we call it the $P_{UCA}$ Framework in this paper. As a continued development, another framework focusing on the prioritisation of requirements is introduced. It is called the $P_{REQ}$ Framework in this paper. The $P_{REQ}$ Framework is the main contribution of this paper. It utilizes the priorities of UCAs, which are the output of the $P_{UCA}$ Framework.} 
This makes the framework applicable at the early stage of the development when STPA is started, rather than waiting until all the requirements have been identified at a later stage of the development. Besides the priorities of UCAs, the requirement ranking process also considers inputs from domain experts, which are then calculated using Monte-Carlo Simulation (MCS) to minimise the subjectivity of the ranking. The calculation is supported by an automation toolchain, and the results can be automatically mapped in a dynamic scaling matrix. This allows the framework to be scalable to high-level systems at the initial conceptualisation of the development and then to more complex systems as the development progresses. The matrix provides a clear visualisation of the ranked sets of requirements to help with further decision-making. The framework was applied to a real project aiming at improving the safe operations of Electric Vertical Take-off and Landing (eVTOL), which is a new development domain that requires continuous identification, prioritisation, and management of requirements. The proposed prioritisation framework has significantly contributed to the decision-making processes throughout the development.
\section{Literature Review}

Prioritisation of requirements has been extensively studied across numerous publications, largely because handling thousands of outputs can strain resource allocation and impede decision-making during system development. Common techniques, such as the AHP, the MoSCoW method, and Cumulative Voting, help stakeholders assess and rank requirements according to value, risk, cost, and detectability. However, these methods often suffer from subjectivity and scalability issues when applied to large data sets, as well as increased complexity and time demands due to a lack of automation tools.

The \cite{tasneem2025enhancing} reviewed 76 primary studies from five reputable electronic databases to examine how different prioritisation techniques are used in Agile Software Development (ASD). Their findings underscore the dynamic nature of ASD and the critical need for scalable, efficient, and automated approaches to maintain alignment between rapidly evolving requirements and stakeholder priorities.

Within the STPA framework, the challenge of prioritising requirements is even more pronounced due to the voluminous outputs STPA can generate. Presenting these requirements to end-to-end stakeholders often leads to duplicated or non-actionable items, which undermines both the efficiency and the quality of the analysis. Moreover, poorly organized results can make communicating safety insights to stakeholders exceedingly difficult \cite{harkleroad2013review}.



\noindent In safety-critical projects, stakeholders increasingly demand demonstrable evidence that residual risk has been reduced to an acceptable level; ISO/IEC\,31010 formalises this ethos by defining risk as “the effect of uncertainty on objectives” and by listing analytical methods for making that uncertainty explicit. Among the 31 quantitative techniques listed in its Annex\,B, \emph{Monte-Carlo simulation}, recommended for evaluating the range and frequency of possible outcomes, provides the probabilistic foundation needed to rank Unsafe Control Actions (UCAs) by expected loss and thus to prioritise safety requirements within the STPA framework, \cite{ISO31010,Pidd2009,olaru2014monte}.
 
In safety-related projects, Stakeholders no longer tolerate uncertainty in safety's output. ISO/IEC 31010 explicitly frames risk as the effect of uncertainty, and Monte Carlo Simulation was listed as one of the recommended techniques for risk assessment. 

MCS is a computational technique originally designed to answer "what-if" scenarios \cite{Pidd2009,olaru2014monte}. It enables the exploration of multiple potential outcomes by running thousands or even millions of simulations, where various input parameters are randomly varied within a predefined range\cite{saltelli2008}. \Gls{mcs} is widely used in numerous disciplines, particularly in risk assessment and decision-making \cite{Boyle1977,fu1995sensitivity}, as it allows for quantifying uncertainty and evaluating different scenarios under changing conditions. When integrated with STPA, \gls{mcs} helps to reduce, if not entirely eliminate, the subjectivity and variability in expert inputs \cite{GoodwinWright2014}, especially when assessing risk severity and likelihood. Instead of relying purely on human judgment, \gls{mcs} leverages computational algorithms to analyse risk factors.



Reducing uncertainty in expert judgments and minimizing the time required to apply existing frameworks motivates the pursuit of improvements in prioritising results within STPA for risk prediction. This paper presents the integration of \gls{mcs} into STPA and proposes an approach to address these challenges. Involving STPA analysts in the process has led to the development of a methodology that offers a more objective means to prioritise requirements and manage them effectively, thereby facilitating the integration and implementation of urgent requirements.
\section{Methodology}
\begin{figure*}[tb]
    \centering
    \includegraphics[width=0.9\linewidth]{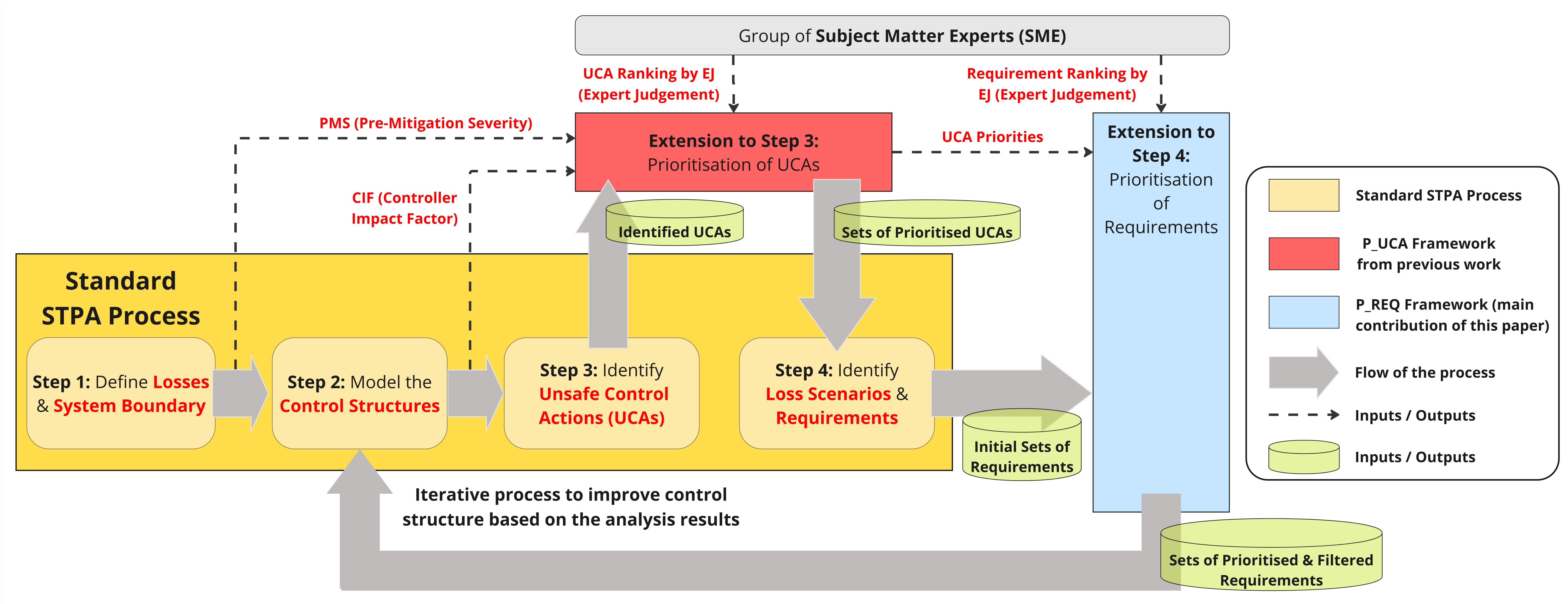}
    \caption{Overall Flowchart of integrating the Standard STPA and the Prioritisation Process}
    \label{fig:Flowchart}
\end{figure*}

An overall flowchart of the whole process is illustrated in Figure \ref{fig:Flowchart}. It includes both the standard STPA process (i.e., the areas in yellow), \textcolor{black}{its extension for the prioritisation of UCAs (i.e., the areas in red), and the extension steps for the prioritisation of requirements (i.e., the areas in blue).} 

\begin{figure}[tb]
    \centering
    \includegraphics[width=\linewidth]{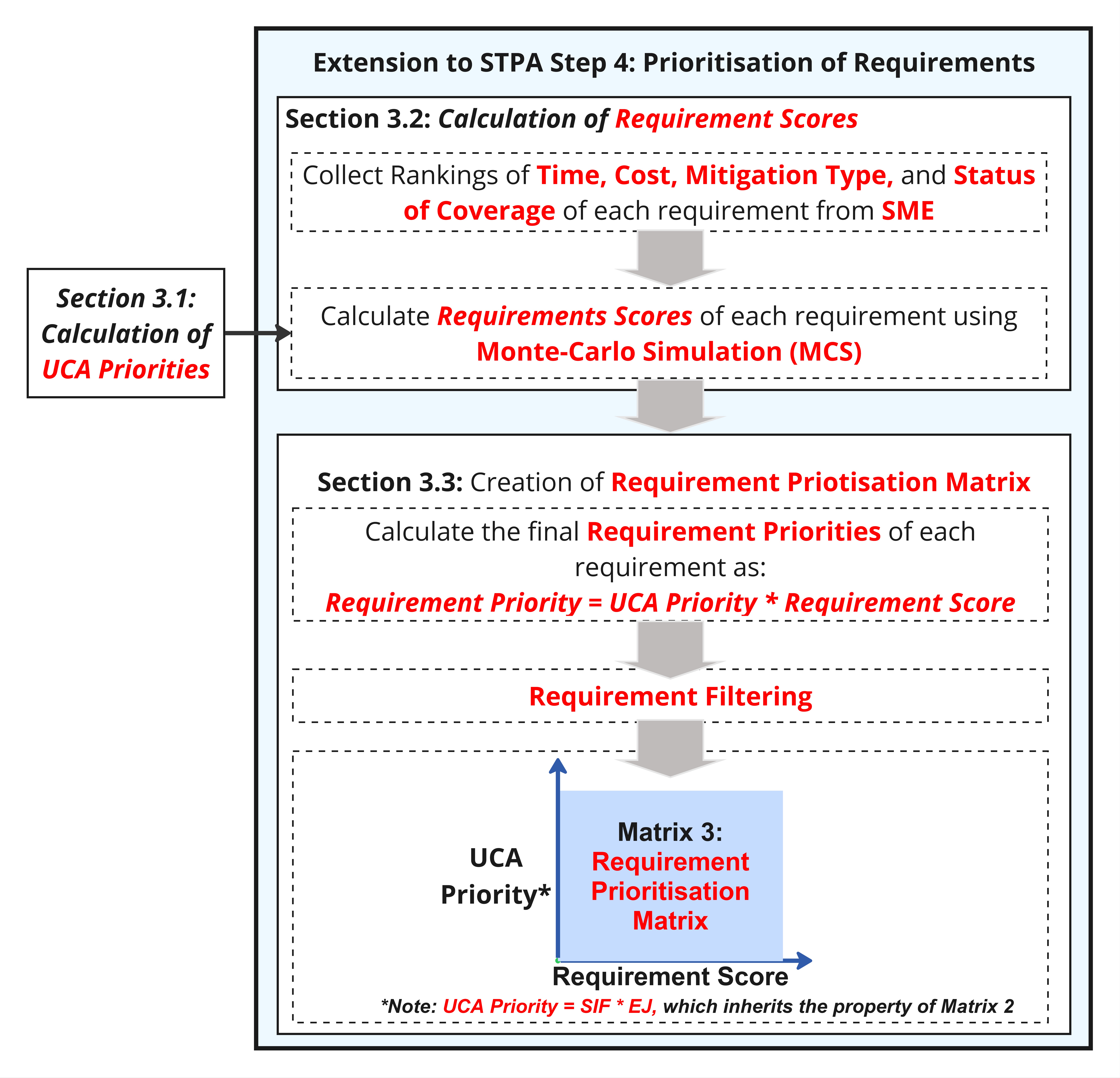}
    \caption{Detailed Flowchart of Requirement Prioritisation Process}
    \label{fig:Req Pri Flowchart}
\end{figure}

A more detailed flowchart of the Extension to Step 4 is illustrated in Figure.\ref{fig:Req Pri Flowchart}. To enable the prioritisation of requirements, two parameters are needed, called the UCA Priority and the Requirement Score. The UCA Priority, as it is named, indicates the priority of the UCA, ranked from P1 (highest priority) to P5 (lowest priority). It is the final output of the \textcolor{black}{$P_{UCA}$ Framework}


\textcolor{black}{It is important to note that the SME involved in the UCA ranking and the requirement ranking can be of the same group of people.} Once the requirements are prioritised, they are then further filtered so that the sets of unique and prioritised requirements are generated. The process of the three parts will be elaborated in the following subsections as follow: 
\begin{itemize}
    \item Calculation of UCA Priorities.
    \item Calculation of Requirement Scores.
    \item Creation of Requirement Prioritisation Matrix.
\end{itemize}

\subsection{Calculation of UCA Priorities}
\begin{figure*}[tb]
    \centering
    \includegraphics[width=0.8 \linewidth]{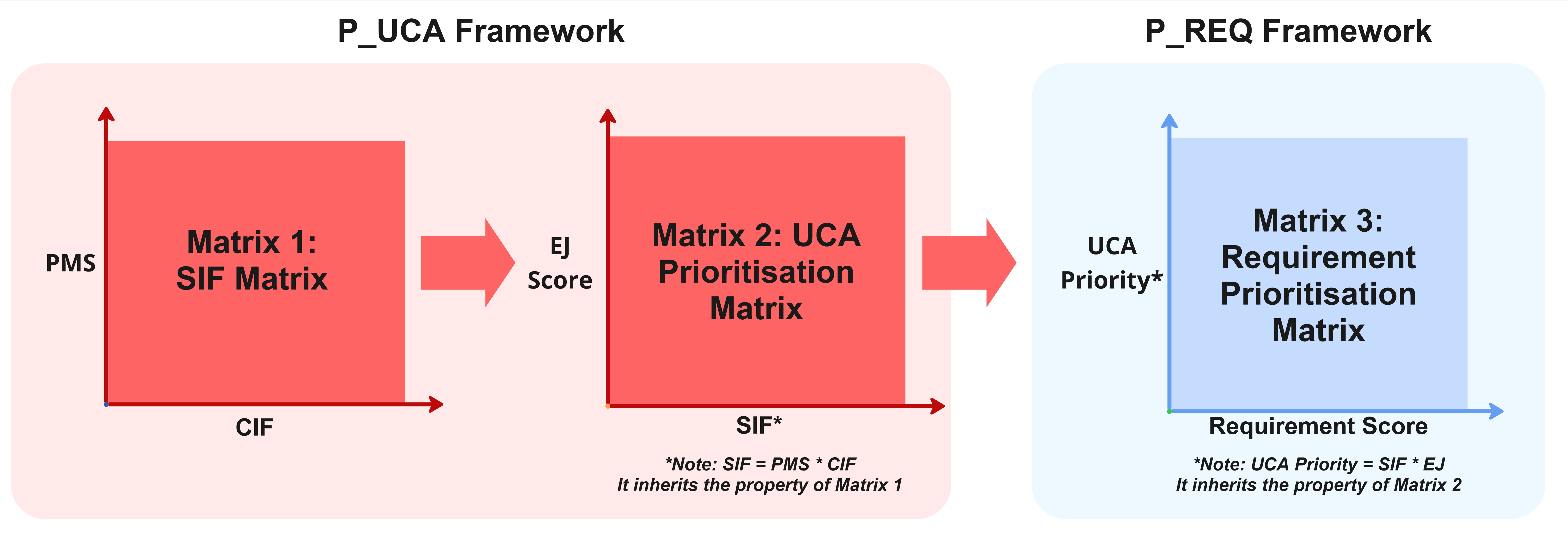}
    \caption{The Matrices generated from both Frameworks and their relationships}
    \label{fig:Matrix}
\end{figure*}

The first parameter of requirement prioritisation is the Priority of the UCA, which we call $P_{UCA}$ in this paper for simplicity. $P_{UCA}$ is the output of the $P_{UCA}$ Framework initially published in our previous work \cite{UCAPrio}. The calculation of the UCA priorities considers the inputs from both STPA experts and SME.

\textcolor{black}{There are several parameters introduced in the $P_{UCA}$ Framework, including Pre-Mitigation Severity (PMS), Controller Impact Factor (CIF), Severity-Impact Factor (SIF), and Expert Judgement (EJ) Score. Details of these parameters can be found in \cite{elbadaoui2025structured}. As the output of the $P_{UCA}$ Framework, two matrices were generated, illustrated as Matrix 1 and Matrix 2 in Figure \ref{fig:Matrix}. The parameter SIF, i.e., the X axis of the Matrix 2, inherits the property of Matrix 1 as the product of the PMS and CIF. Together with the EJ Score (i.e., the Y axis of the Matrix 2), they provide a visualisation of UCA ranking from both the safety analysts and the SME.}

\subsection{Calculation of Requirement Scores}
The second parameter required to calculate the priorities of requirements is called the Requirement Score (RS). This involves collecting the rankings from SME and calculating the EJ scores based on the rankings to generate the final requirement scores using MCS. Both are elaborated in the following subsections.

\subsubsection{Collecting EJ Scores from the SME}
Similar to the EJ Ranking process for UCA Prioritisation, the RS is also calculated based on several factors that need to be scored by the SME. Compared to the EJ Ranking for UCA Prioritisation, the RS is based on the following four factors: 1) The \textbf{Time} required to implement the requirement; 2) The \textbf{Cost} of implementing the requirement; 3) The Type of the requirements; and 4) Whether the Requirement has been \textbf{covered} by pre-existing regulations. Each factor is assigned a set of thresholds that indicate different levels of intensity. These intensities are designed to be clear and unique to each factor \cite{GoodwinWright2014} and provide SMEs guidance during their assessment. The final score of each requirement is calculated based on the ranked intensities of each factor. A requirement with a higher score represents a higher priority.

Table \ref{Tab:Requirement Score Factors} summarizes the four factors with intensities. The \textbf{Time} factor was ranked based on the amount of effort required to implement the requirement, either requiring significant, moderate, or minor effort. A requirement ranked \textbf{Minor Time} has a higher score than one ranked \textbf{Significant Time}. Similar to the \textbf{Time} factor, the \textbf{Cost} factor was also ranked based on the amount of cost required to implement the requirement, which is ranked as either \textbf{High Cost}, \textbf{Medium Cost}, or \textbf{Low Cost}. A requirement ranked \textbf{Low Cost} has a higher score than the one ranked \textbf{High Cost}. The \textbf{Mitigation Type} factor indicates the risk reduction strategies in safety-critical systems. These mitigation types categorize how a safety-related hazard, or Causal Factors (CF), in the context of STPA, is addressed. These categorizations have often been used in safety and hazard analysis literature, such as \cite{leveson2016engineering}. Based on the categorizations, there are five types of mitigation: \textbf{Type A)} The mitigation involves selecting a fundamentally different design, material, or system architecture to eliminate the CF; \textbf{Type B)} The mitigation consists of modifying the existing design to reduce the severity or likelihood of the CF without eliminating it; \textbf{Type C)} CF is reduced through active features such as automatic shutdowns, barriers, interlocks, or emergency systems; \textbf{Type D)} CF is reduced by developing detection mechanisms and inform operators or systems of a developing hazard; \textbf{Type E)} CF is reduced by involving training, policies, signage, Personal Protective Equipment (PPE), or standard operating procedures (SOPs). The scores of \textbf{Type A }requirements to \textbf{Type E} requirements decrease - i.e., a \textbf{Type A} requirement has the highest score and a \textbf{Type B} requirement has the lowest score. Because implementing the \textbf{Type A} requirement fully eliminates the CF. The last factor that needs to be considered is whether the requirement has been \textbf{covered} by pre-existing regulations or not. The intensities for this factor are binary, where a \textbf{1} represents that the requirement has not been covered by pre-existing regulation and the CF is likely to occur, and a \textbf{0} represents that the requirement has already been covered by pre-existing regulations and the CF is unlikely to occur. A requirement ranked as \textbf{1} has a higher score than the one ranked as \textbf{0} because it represents a newly found requirement that has been overlooked in the industry, which needs to be prioritised.

\begin{table}[t]
\caption{Factors of Requirement Score with Intensities}
\label{Tab:Requirement Score Factors}
\centering
\begin{tabular}{|p{0.1\textwidth}|p{0.35\textwidth}|} \hline  
Parameters & Descriptions \\ \hline 
Time &
\textbf{Significant:} Implementing the Requirement would require significant effort. \newline
\textbf{Moderate:} Implementing the Requirement would require moderate effort. \newline
\textbf{Minor:} Implementing the Requirement would require minor effort. \\
\hline 

Cost & 
\textbf{High:} The cost involved in implementing the Requirement would be high. \newline
\textbf{Medium:} The cost involved in implementing the Requirement would be medium. \newline
\textbf{Low:} The cost involved in implementing the Requirement would be low. \\ \hline 

Mitigation Type &
\textbf{Type A:} The Requirement is implemented by selecting a major design or material alternative to eliminate the CF. \newline
\textbf{Type B: }The Requirement is implemented by a minor design change that reduces the severity and/or the probability of the CF. \newline
\textbf{Type C: }The Requirement is implemented by using engineered features or devices to actively interrupt the mishap sequence and reduce the risk of the mishap. \newline
\textbf{Type D: }The Requirement is implemented by including detection and warning systems to alert personnel to the CF. \newline
\textbf{Type E: }The Requirement is implemented by incorporating signage, procedures, training, and PPE to the personnel. \\ \hline 

Covered by Regulations? & 
\textbf{1:}The Requirement has not been covered by pre-existing regulations, and the CF is likely to occur. \newline
\textbf{0:} The Requirement has already been covered by pre-existing regulations, and the CF is unlikely to occur. \\ \hline
\end{tabular}
\end{table}

\subsubsection{Calculating EJ Scores using MCS}
By integrating SME insight, we can quantify how each STPA requirement affects cost, time, and novelty of the requirements. Within STPA, SME feedback enriches Control Structure diagrams, clarifies UCAs, and yields comprehensive, realistic safety constraints. This multi-perspective view elevates the quality of the study and uncovers risks that might be overlooked by a single analyst. The trade-off is the uncertainty that every judgment reflects the expert’s perspective. 

To address the need to reduce uncertainty, we will apply an MCS approach. This method aims to prioritize the UCAs by making expert inputs more objective. By doing so, we can minimize the impact of subjectivity on our analysis results and clearly identify the most relevant requirements.

\begin{figure*}[tb]
    \centering
    \includegraphics[width=1\linewidth]{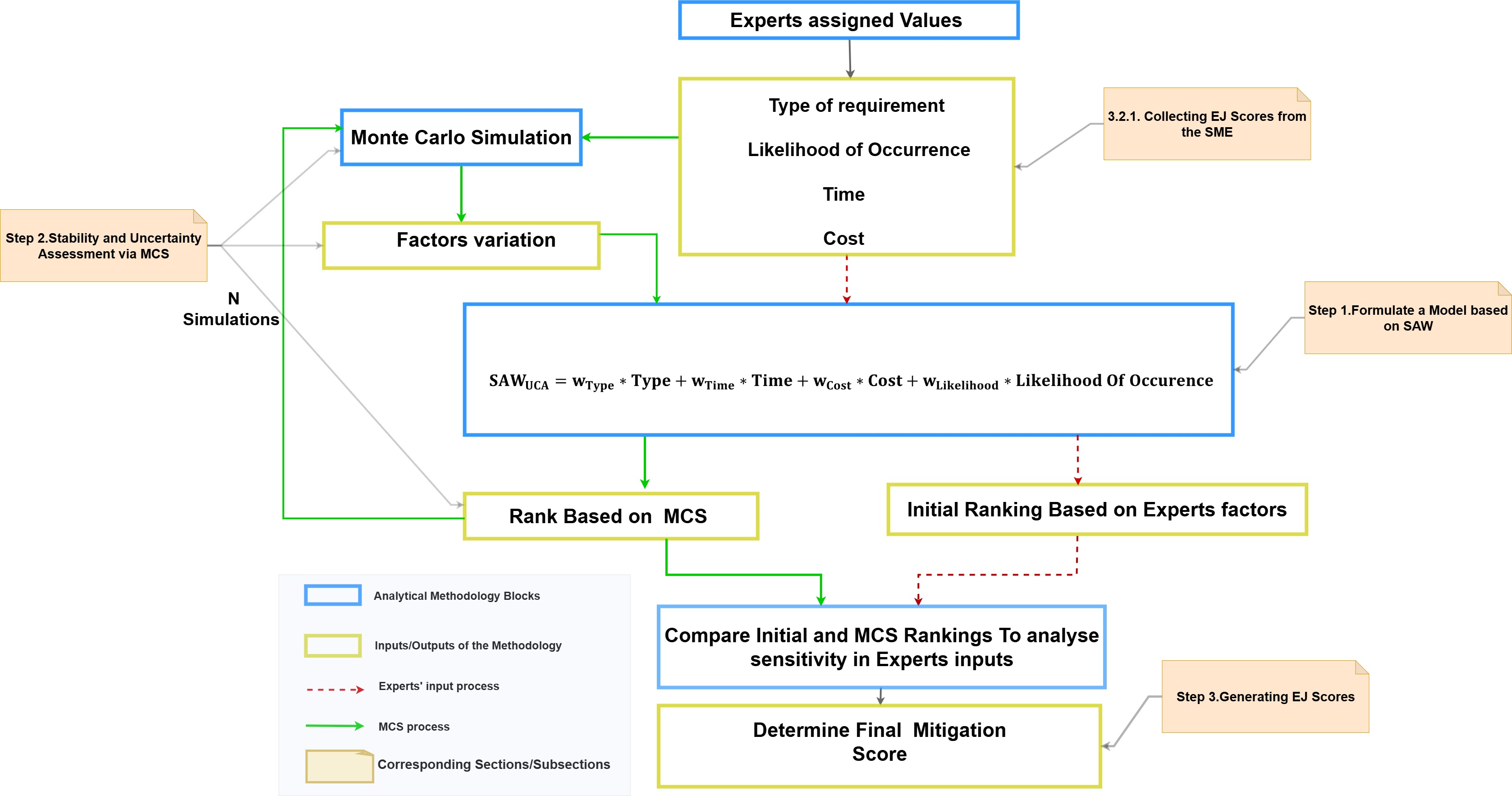}
    \caption{Overview of the Requirements Score calculation Process and Corresponding subsections/ Steps in the paper}
    \label{fig:RS calculation}
\end{figure*}

Figure \ref{fig:RS calculation} illustrates the methodology used to calculate the RS. The following paragraph outlines the detailed steps for computing the RS based on the MCS: 
\begin{itemize}
    \item \textbf{Step 1.Formulate a Model based on SAW}
    \item \textbf{Step 2.Stability and Uncertainty Assessment via MCS}
    \item \textbf{Step 3.Generating EJ Scores}
\end{itemize}

\subsection*{\textbf{Step 1.Formulate a Model based on SAW}}
The objective of this step is to convert the four evaluation factors, type, time, cost, and novelty, into a single numerical score that reflects the overall relevance and impact of the requirement.
The weighted scores are aggregated using the Simple Additive Weighting (SAW) method to produce an initial ranking of requirements based on SME inputs.


\begin{equation}
\label{SAW}
\mathrm{SAW}_{\mathrm{UCA}} = w_{\mathrm{Type}} \cdot \mathrm{Type} + w_{\mathrm{Time}} \cdot \mathrm{Time} + w_{\mathrm{Cost}} \cdot \mathrm{Cost} + w_{\mathrm{Likelihood}} \cdot \mathrm{L_{occ}}
\end{equation}

\textbf{Where:} \\

$w_{Type}$ = 0.4, \quad $ w_{Likelihood}$ = 0.3, 

$w_{Time}$ = 0.15, \quad $w_{Cost}$ = 0.15


Next, we \emph{assign weights} to each criterion so that the final score reflects the relative importance of the requirement, in terms of its novelty and the amount of resource and effort needed to address the requirement. In practice:

   \begin{itemize}
      \item \emph{Requirement Type:} \textbf{Type A} requirements receive the highest weight where: {$w_{Type}$ = 0.4}.
      \item \emph{Regulatory Gap: }Requirements not \textbf{covered} by pre-existing regulations score highly for novelty, {$w_{Likelihood}$ = 0.3}.
      \item \emph{Implementation Cost:} Lower \textbf{financial} outlay { $w_{Cost}$ = 0.15}.
      \item \emph{Implementation Time:} Requirements demanding minimal \textbf{time} to be implemented  {$w_{Time}$ = 0.15}.
    \end{itemize}

\subsection*{\textbf{Step 2.Stability and Uncertainty Assessment via MCS}}

After establishing the initial ranking based on experts, we examine its stability by varying the uncertainty of each factor through the MCS. The procedure consists of these steps:

\subsection*{Step 2.1 Parametrise the uncertainty}
For every decision factor, we elicit a \emph{triangular} distribution  
\emph{Tri}(a,c,b), where \(a\) is the lower limit, \(b\) the upper limit, and \(c\) the mode (the expert’s estimation).

\subsection*{Step 2.2 Sensitivity Analysis}

\begin{itemize}
  \item Fix all factors at their modal values \(c\).
  \item   For each factor in turn, set its value to \(a\) and recompute the SAW score. 
  \item   Repeat with the factor set to \(b\).  
\end{itemize}

A significant shift in rank reveals high influence and highlights the factor for further investigation. If the modal assessment for \textbf{Time} is \(c = 2\) (i.e., Moderate), we may bracket the uncertainty by \(a = 1\) (i.e., Minor) and \(b = 3\) (i.e., Significant). Re-evaluating the SAW score with \textbf{Time} fixed at \(a\) and \(b\) highlights the sensitivity of the ranking to schedule assumptions.

Monte Carlo Simulation Algorithm:
\begin{itemize}
  \item \textbf{Iterations:} Run the simulation for \(N\) iterations (e.g., \(N=1000\)).  
  \item \textbf{Loop.} For iteration \(i = 1,\dots,N\):  
        \begin{enumerate}
            \item Vary each factor by \( \pm10\% \) over \(N\) simulations to measure how small input changes affect rankings.
            \item recompute the corresponding SAW score and store the resulting rank.  
        \end{enumerate}
\end{itemize}
The ensemble of ranks provides confidence intervals for each requirement and highlights those with statistically uncertain rankings.

\subsection*{\textbf{Step 3.Generating EJ Scores}}
\subsection*{Step 3.1 Calculating the Average Rank}

After generating the MCS data, we compute the average rank of each requirement to assess its typical priority:

Let \(N\) be the total number of simulations, and let \(\mathrm{Rank}_{s,j}\) denote the rank of requirement \(j\) in simulation \(s\).  Then the \emph{average rank} of requirement \(j\) is
\begin{equation}
 \label{averge rank}   
\overline{\mathrm{Rank}}_{j}
= \frac{1}{N} \sum_{s=1}^{N} \mathrm{Rank}_{s,j}
\end{equation}

This value represents the central tendency of the requirement’s ordinal position across all simulations.

\subsection*{Step 3.2 Computing Rank Variability}

To measure stability, we calculate the standard deviation of the ranks for each requirement:

\begin{equation}
    \label{standard deviation}
\sigma_{j}
= \sqrt{\frac{1}{N} \sum_{s=1}^{N} \bigl(\mathrm{Rank}_{s,j} - \overline{\mathrm{Rank}}_{j}\bigr)^{2}}
\end{equation} 
A small \(\sigma_{j}\) indicates that the requirement’s ranking is consistent, while a larger \(\sigma_{j}\) reveals greater sensitivity to input uncertainty.

\subsection*{Step 3.3 Calculating the $\emph{Requirement}_{Scores}$}
Finally, we combine centrality and variability into a single metric, the $\emph{Requirement}_{Scores}$ for each requirement:

\begin{equation}
\label{RS}
\mathrm{Requirement\_Score}_{j}
= \overline{\mathrm{Rank}}_{j} + \sigma_{j}
\end{equation}
Lower $\emph{Requirement}_{Scores}$ denote requirements that are both highly ranked and stable under uncertainty.

\subsection*{Step 3.4 Confidence-Interval}

For every UCA a CI is computed to ensure the robustness of the final rank. The upper limit of the interval is applied as follows: 

\begin{equation}
\text{CI}^{\uparrow}_{95}=
\overline{\mathrm{Rank}}_{j}+
1.96\,
\frac{\sigma_{\text{j}}}{\sqrt{\text{num\_simulations}}},
\label{eq:upper_CI}
\end{equation}

The MCS thus yields, for each requirement, a $\emph{Requirement}_{Scores}$ that integrates its average priority and rank dispersion. In the next section, we will merge these scores with the UCAs Priority within the prioritisation matrix, to give our final rank




\subsection{Creation of Requirement Prioritisation Matrix}
\subsubsection{Calculation of Final Requirement Priority}
Once both $P_{UCA}$ and RS are identified for each requirement, the final requirement priority, which we call $P_{Requirement}$, can be calculated. It is important to note that a requirement cannot be prioritised solely based on the description on its own (i.e., the Requirement Score). It is also necessary to identify the priorities of higher-level hazards that the requirement links to. In the context of STPA, we need to consider the priorities of the UCAs as well as the Losses that the requirement is trying to address. The $P_{UCA}$ parameter covers both the priorities of the UCAs and the Losses (i.e., based on the PMS). The final requirement priority is calculated as the product of the $P_{UCA}$ and Requirement Score as shown in equation (\ref{eq2}). In this case, each requirement inherits the priority of the UCA and Losses that it links to. 

\begin{equation}
\label{eq2}
    P_{Requirement} = P_{UCA} * RS
\end{equation}

\subsubsection{Filtering requirements}

Given the voluminous results that can be generated from STPA, even after prioritisation, implementing the highest-level requirements can be challenging. In this real-world case study, where actors of requirements (e.g., CAA) are expected to implement these requirements to enhance regulation, providing raw results is not feasible due to the presence of duplicated outputs. It is important to note that the duplicated requirements from STPA may have been derived from different CFs and UCAs. To maintain the traceability and to avoid the duplication of requirements, an automatic requirement filtering mechanism is developed to identify and merge duplicate requirements. The filtered requirements are automatically populated to an Excel file, where each row corresponds to a unique requirement and includes:

\begin{itemize}
\item \textbf{Req ID}: the identifier linked to its underlying UCA, with the cell background colour automatically generated to reflect its importance level.
  \item \textbf{UCA Description}: a description of the UCA.  
  \item \textbf{Causal Factor(s)}: the contributing factors that give rise to the UCA.  
  \item \textbf{Req Description}: the textual safety requirement intended to mitigate the UCA.  
  \item \textbf{Priority}: the assigned priority label (e.g.\ \texttt{ReqP1}, \texttt{ReqP2}, ...).
\end{itemize}

The automatic colour coding assists stakeholders in identifying high criticality \textit{REQUIREMENT} for implementation. This requirements filtering is performed before mapping results into the matrix. It provides insight into scenarios that will be mitigated by linking each requirement to its causal factors and associated UCA.
In case where different priorities are assigned to the same requirement, the priority column highlights the corresponding priority.

\subsubsection{Mapping Requirements to the final Matrix}

For better visualisation and communication of the $P_{Requirement}$, each requirement, in the form of the unique Requirement ID, can be mapped to the final matrix - i.e., the Matrix 3 of Figure \ref{fig:Matrix}. The matrix is designed in a way that the x-axis represents the calculated RS of the requirement, and the y-axis represents the $P_{UCA}$ of the UCA that the requirement links to.

After filtering the requirements and calculating all factors needed to generate the final rank of requirements, we opted for the choice of a 5*5 Matrix to enable a clear communication of the result and direct visualisation of the high-impact requirements. Each zone of the 5*5 Matrix has a unique colour code. The boundaries of the zones are dynamically scaled based on the data set provided.

Scale \texttt{UCA\_{P}} and \texttt{RS} to a 5 x 5 grid
\[
\text{UCA}_{\text{P},Scaled} = \left\lfloor \left( \frac{\text{P}_{\text{UCA}}}{\max\left(\text{P}_{\text{UCA}}\right)} \right) \times 4 \right\rfloor
\]
\[
\text{RS}_{Scaled} = \left\lfloor \left( \frac{\text{RS}}{\max\left(\text{RS}\right)} \right) \times 4 \right\rfloor
\]

where: \( \text{UCA}_{\text{P},Scaled} \) and \(\text{RS}_{Scaled} \) are the scaled values for \( \text{P}_{\text{UCA}} \) and \( \text{RS} \), respectively, mapped to a 5-level scale (0 to 4).

\begin{figure}
    \centering
    \includegraphics[width=1\linewidth]{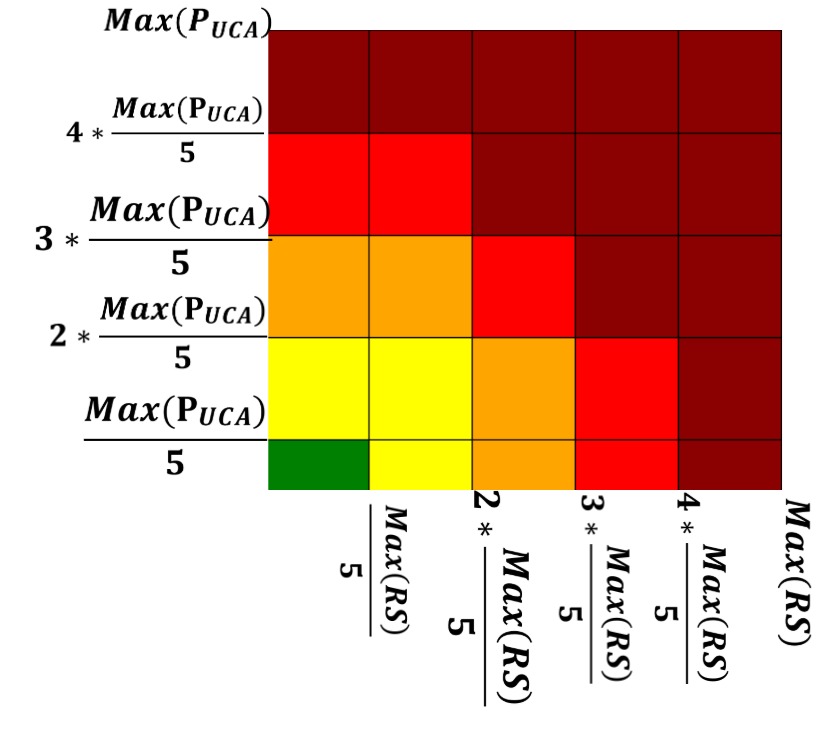}
    \caption{Overview of the dynamic scaling of data in 5*5 Matrix}
    \label{fig:Dynamic}
\end{figure}

We set up dynamic scaling by finding the maximum values for $\text{P}_{\text{UCA}}$ and \texttt{RS}. We define a colour gradient to represent varying levels of risk severity, ensuring the matrix adapts to the dataset’s range without requiring fixed grid dimensions. Figure \ref{fig:Dynamic} shows a detailed description of how range levels are created based on input data.

\section{Case study and Results}
\begin{table*}[tb]
\centering
\caption{Requirements along with their corresponding factors based on SME input}
\label{tab:req-score}
\begin{tabular}{ccccc}
\toprule
\textbf{Req ID} & \textbf{Time}          & \textbf{Cost}        & \textbf{Type} & \textbf{Likelihood} \\ \midrule
UCA(Ph2)- 7.5.2-RQ.5    & Moderate effort    & Medium (30–60\%)  & Type C & 1 \\ 
UCA(Ph2)- 7.5.3-RQ.2    & Moderate effort    & Medium (30–60\%)  & Type C & 1 \\ 
UCA(Ph0.1)-13.5.2-RQ1   & Minor effort       & Low (below 30\%)  & Type A & 1 \\ 
UCA(Ph0.1)-14.5.1-RQ1   & Minor effort       & Low (below 30\%)  & Type A & 1 \\ 
UCA(Ph0.1)-15.5.1-RQ1   & Moderate effort    & Low (below 30\%)  & Type A & 1 \\ 
UCA(Ph1)-18.2.1-RQ1     & Minor effort       & Low (below 30\%)  & Type E & 0 \\ 
UCA(Ph0.2)-33.7.2-RQ2   & Minor effort       & Low (below 30\%)  & Type E & 0 \\ 
UCA(Ph1)-18.5.1-RQ2     & Minor effort       & Low (below 30\%)  & Type E & 0 \\ 
UCA(Ph1)-18.2.2-RQ1     & Minor effort       & Low (below 30\%)  & Type E & 0 \\ 
UCA(Ph1)-18.2.2-RQ5     & Minor effort       & Low (below 30\%)  & Type E & 0 \\ 
UCA(Ph0.1)-34.1.1-RQ2   & Minor effort       & Low (below 30\%)  & Type E & 0 \\ 
UCA(Ph0.2)-33.1.2-RQ2   & Minor effort       & Low (below 30\%)  & Type E & 0 \\
UCA(Ph0.2)-10.6.1-RQ2 & Minor effort & Low(below 30\%) & Type D & 1 \\
UCA(Ph0.1)-17.1.2-RQ1 & Minor effort & Low(below 30\%) & Type B & 1 \\
UCA(Ph0.1)-49.5.1-RQ4 & Moderate effort & Low(below 30\%) & Type B & 1 \\

\bottomrule
\end{tabular}
\end{table*}

\begin{figure*}[tb]
    \centering
    \includegraphics[width=\linewidth]{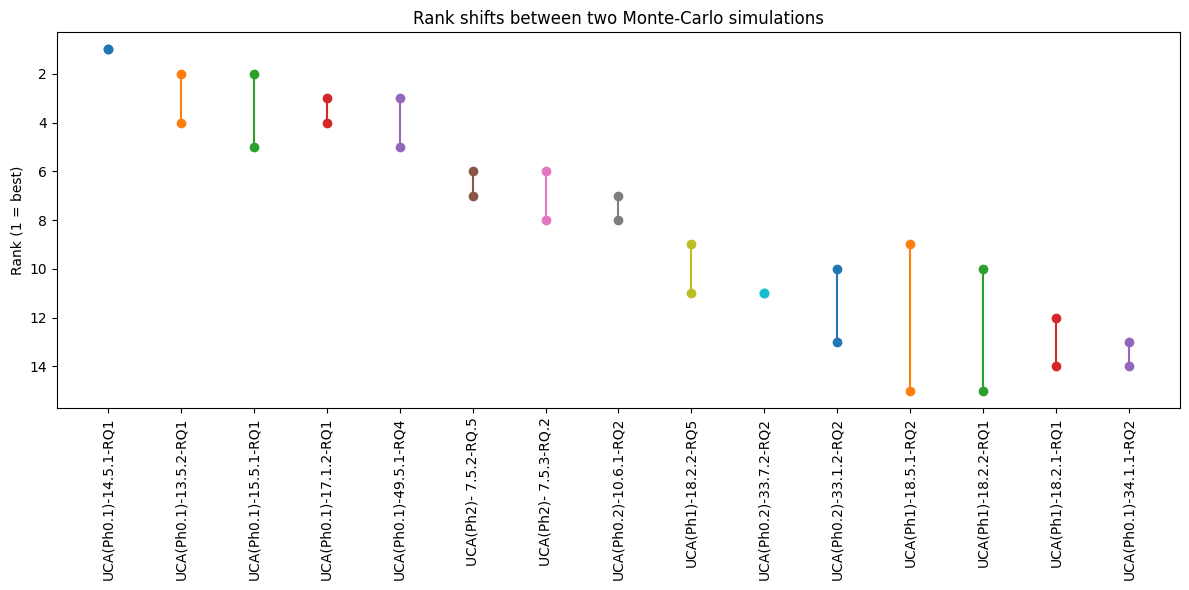}
    \caption{Rank-shift diagram of requirement priorities between two independent Monte-Carlo simulations}
    \label{fig: MCS}
\end{figure*}

\begin{figure*}[tb]
    \centering
    \includegraphics[width=0.8\linewidth]{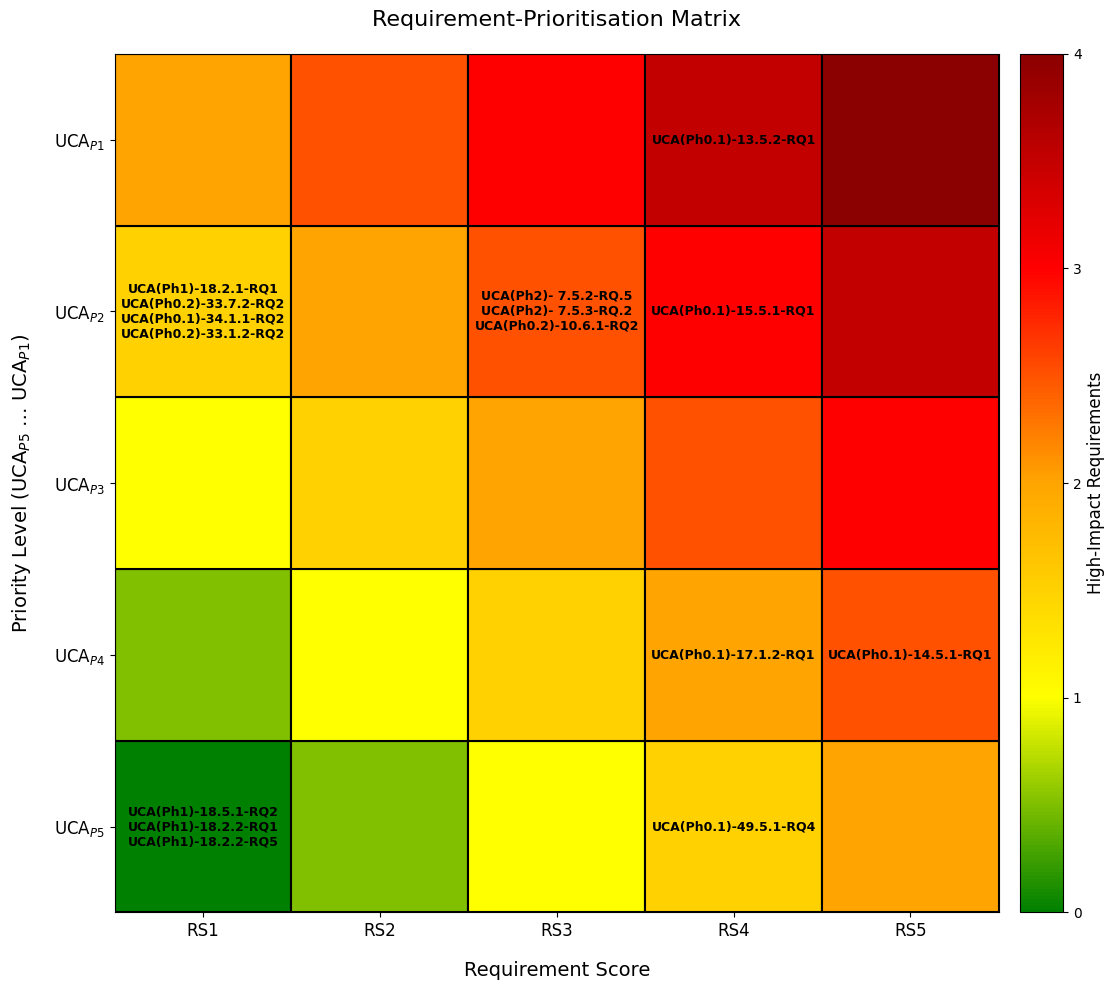}
    \caption{Prioritisation Matrix of Requirements, where $\text{{UCA}}_{Pn}$ refers to the priority level of the UCAs, and  $\text{RS}_n$ refers to the requirement score range.}
    \label{fig: Prioritisation Matrix of Requirements}
\end{figure*}

\begin{figure*}[p]
    \centering
    \includegraphics[width=\linewidth]{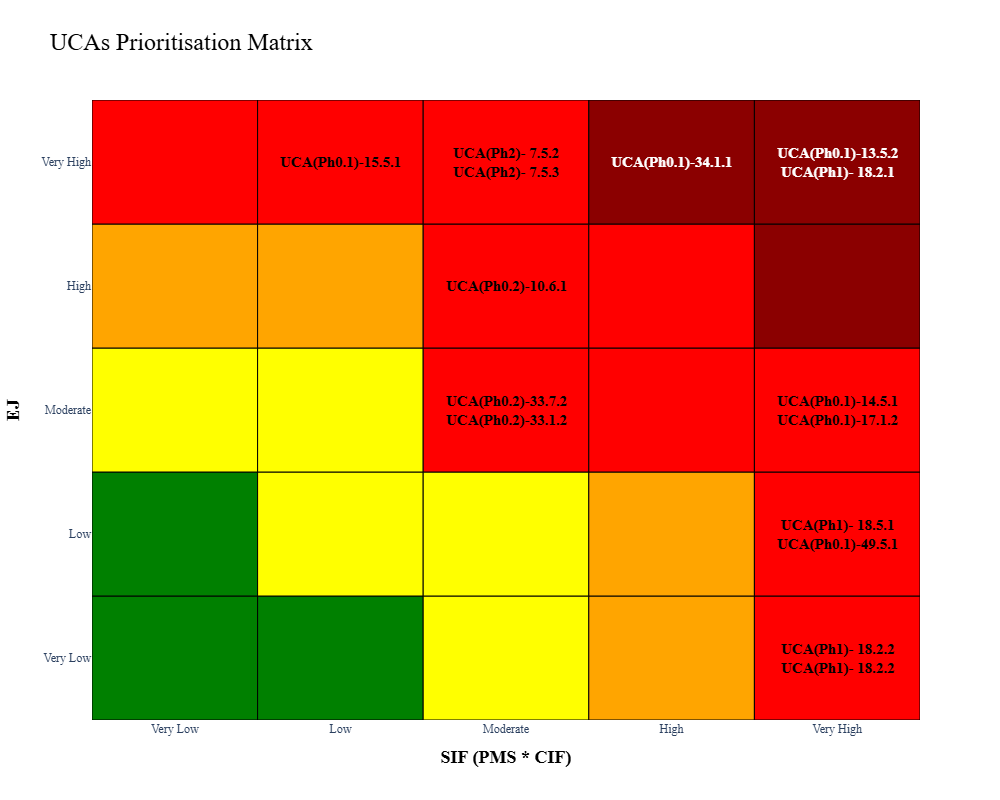}
    \caption{Prioritisation Matrix of UCAs}
    \label{fig: Prioritisation Matrix of UCAs}
\end{figure*}

This work is an extension of previous research on the prioritisation of UCAs \cite{UCAPrio}, conducted in the context of collaboration between WMG and the CAA. In which the Risk Sub-group of the eVTOL Safety Leadership Group (eVSLG) used the STPA results to highlight regulatory gaps and address them.
The collaboration resulted in Civil Aviation Publication that support the future integration of electric vertical take-off and landing (eVTOL) aircraft into UK airspace. The study identifies over 50 high-priority areas to develop regulations \cite{caa_stpa_evtol}.
In this section, we present the results and findings for each proposed concept.

\noindent In our eVTOL STPA study, we divided the analysis into five phases, each with its own control structure, to keep the process both manageable and readable:
\begin{itemize}
  \item \textbf{Phase 0.1} (\texttt{Ph0.1}): Regulatory preparation  
  \item \textbf{Phase 0.2} (\texttt{Ph0.2}): Operational preparation  
  \item \textbf{Phase 1}   (\texttt{Ph1}): Take‐off  
  \item \textbf{Phase 2}   (\texttt{Ph2}): Cruise  
  \item \textbf{Phase 3}   (\texttt{Ph3}): Descent and landing  
\end{itemize}

\noindent By prefixing each requirement with its UCA and phase identifier, we ensure that:
\begin{itemize}
  \item The context of each requirement is immediately clear.
  \item Requirements remain unique and traceable across phases.
  
\end{itemize}

\subsection{Calculation of UCA Priorities}

Table~\ref{tab:uca-p} presents the prioritisation results of UCAs. Each UCA was assigned a score based on its \textit{SIF~$\times$~EJ} value and mapped to a corresponding priority level. The EJ score is derived from a sensitivity analysis, with lower EJ values corresponding to higher criticality. To incorporate EJ into the SIF calculation, we first invert each EJ and then multiply by the SIF.

\textcolor{black}{Each Req ID in Table \ref{tab:uca-p} encodes three pieces of information, including the UCA to which this requirement is derived from, the phase of Control Structure in which the UCA was identified from, and a sequential requirement number within that phase. For example, the Req ID, UCA(Ph2)\,--\,7.5.2-RQ.5 in the same table denotes the fifth requirement identified for UCA(Ph2)-7.5.2, which was identified from the Phase 2 Control Structure.} \textcolor{black}{It is important to note that strategically, this stage of the analysis only focuses on the identification of requirements of P1 and P2 UCAs, which have higher priorities and are more critical.}

Figure \ref{fig: Prioritisation Matrix of Requirements} shows the output of the prioritisation concept for the UCAs. Each UCA is positioned within the matrix according to its scaled EJ score and SIF value. The background colour of each cell reflects the criticality level. This visual mapping improves the output communication and decision-making by highlighting which UCAs require the most urgent attention.


\subsection{Calculation of Requirement Scores}
Table \ref{tab:req-score} shows the scores assigned by SMEs to each requirement across the four evaluation factors: Type, Cost, Time, and Likelihood.


Figure \ref{fig: MCS} illustrates the sensitivity of requirement ordering to uncertainty rank shifts between two Monte Carlo simulations. Requirements with short segments present a stable rank and confirm rank robustness based on expert inputs. Conversely, requirements showing shifts of at least five places in the ranking demonstrate a need to refine the data. Colours distinguish individual lines only and do not carry any analytic meaning, and follow Matplotlib’s default colour cycle. 


\subsection{Creation of Requirement Prioritisation Matrix}


Table~\ref{tab:reqFiltring} presents the filtered requirements output as generated in Excel. Each row corresponds to a unique requirement.

All requirements are unique (no duplicates). The automatic colour‐coding helps stakeholders quickly identify high‐criticality requirements for implementation.

In Figure \ref{fig: Prioritisation Matrix of Requirements} each cell shows the requirement IDs whose Monte-Carlo mean score falls into class \texttt{RS1}–\texttt{RS5} on the horizontal axis which represents the commercial score, and requirements whose inverted-EJ* SIF lies in priority band \texttt{P5}–\texttt{P1} on the vertical axis, which reflects the UCA-priority score.  Background colours progress from green (low impact) to dark red (highest impact); the colour-bar on the right quantifies this scale, it ranges from 0 (green) to 4 (dark-red) and allows an immediate visual ranking of the set reviewed. The matrix provides a traceable, one-glance overview of which requirements deliver the greatest safety benefit for the least commercial effort.

The priority levels \texttt{P1}–\texttt{P5} that appear in the Requirement-Prioritisation Matrix are \emph{not} the same bands used in the earlier UCA-prioritisation matrix.  
To avoid confusion, we proceeded as follows:

\begin{enumerate}
  \item From the UCA matrix we retained only those UCAs that had been classified as \texttt{UCA\_{P1}} or \texttt{UCA\_{P2}}.  
  \item For every requirement mitigating one of these UCAs we computed the product SIF* EJ.
  \item The resulting set of scores was then split into \emph{five new quantiles}, generating a fresh 5 × 5 scale for the Requirement Matrix. 
\end{enumerate}

\begin{table}[tb]
\centering
 \caption{Output of the UCA's prioritisation concept: SIF* EJ}
\label{tab:uca-p}
\begin{tabular}{llll}
\toprule
\textbf{Req ID} & \textbf{EJ } & \textbf{SIF} & \textbf{SIF* EJ}  \\
\midrule
UCA(Ph2)\,--\,7.5.2-RQ.5   & 29.79 & 60  & 42.12 \\
UCA(Ph2)\,--\,7.5.3-RQ.2   & 30.03 & 60  & 41.98 \\
UCA(Ph0.1)\,--\,13.5.2-RQ1 &  6.95 & 160 & 148.89 \\
UCA(Ph0.1)\,--\,14.5.1-RQ1 & 98.10 & 160 &   3.04 \\
UCA(Ph0.1)\,--\,15.5.1-RQ1 & 29.82 &  56 &  39.30 \\
UCA(Ph1)\,--\,18.2.1-RQ1   & 29.86 & 140 &  98.20 \\
UCA(Ph0.2)\,--\,33.7.2-RQ2 & 97.91 &  60 &   1.26 \\
UCA(Ph1)\,--\,18.5.1-RQ2   & 162.59 & 140 &   0.00 \\
UCA(Ph1)\,--\,18.2.2-RQ1   & 208.26 & 140 &   0.00 \\
UCA(Ph1)\,--\,18.2.2-RQ5   & 208.26 & 140 &   0.00 \\
UCA(Ph0.1)\,--\,34.1.1-RQ2 & 29.88 & 120 &  84.15 \\
UCA(Ph0.2)\,--\,33.1.2-RQ2 & 98.05 &  60 &   1.17 \\
UCA(Ph0.2)\,--\,10.6.1-RQ2 & 59.27 &  80 &  32.58 \\
UCA(Ph0.1)\,--\,17.1.2-RQ1 & 97.64 & 160 &   3.78 \\
UCA(Ph0.1)\,--\,49.5.1-RQ4 & 162.39 & 160 &   0.00 \\
\bottomrule
\end{tabular}
\end{table}

\section{Discussion}

STPA has been recognised as a recommended method for the analysis of the increasingly complex modern systems to identify safety-critical issues and requirements that are overlooked by traditional methods. However, there is still room to improve the process of STPA, especially in the context of a fast-paced development cycle. A scalable framework to enable the prioritisation of STPA requirements is introduced in this paper. It does not require any change to the standard STPA process, allowing the prioritisation framework to be easily applied by the development team if standard STPA is already part of the procedure. The prioritisation framework utilizes the outputs from all steps of STPA, which includes the PMS from Step 1 (Define Purposes of the Analysis), CIF from Step 2 (Model the Control Structure), UCA Priorities from Step 3 (Identify UCAs), and finally the Requirement Scores from Step 4 (Identify Loss Scenarios and Requirements). This allows the prioritisation framework to be applied right from the start of the development lifecycle, in parallel with the standard STPA. As mentioned in the \cite{UCAPrio}, the UCA Prioritisation framework results in sets of UCAs with different priorities, from P1 (highest priority) to P5 (lowest priority). At this stage, developers can decide whether or not to directly start Step 4 from those top priority UCAs. In the eVTOL Operation case study, UCAs with priorities P1 and P2 were first analysed in Step 4, making the framework more flexible in the agile development environment. This bridges the research gap \textbf{RQ1 }(i.e., How can the requirements identified from the STPA be effectively implemented for safe system operation?).


Bringing SMEs into the ranking process adds significant value because their practical insight helps rate requirements more effectively.  
The other side of the coin, however, is \emph{uncertainty}: experts from different organisations may assign different scores to the same requirement.  
To quantify and mitigate this variability, we adopt an MCS approach.  
By propagating the full range of SME estimates through a large number of simulations, MCS provides an objective ranking that reflects the collective expert view while reducing the influence of individual bias.

Figure \ref{fig: MCS} illustrates how sensitive the requirement ranking is to variations in the input data.  
The diagram compares two independent Monte-Carlo simulations, for each requirement, the vertical distance between its two ranks: a single dot (zero-length bar) means the rank did not change, whereas a longer bar shows how far the requirement moved up or down the list.

\begin{itemize}
  \item Requirements such as \texttt{UCA(Ph0.1)-14.5.1-RQ1} and \texttt{UCA(Ph0.2)-33.7.2-RQ2} appear only as dots and their positions are essentially fixed across simulations.  
  \item Other requirements, for example \texttt{UCA(Ph0.1)-15.5.1-RQ1} and \texttt{UCA(Ph0.1)-17.1.2-RQ1}, show bars only one or two ranks long, indicating minor, acceptable variability.  
  \item By contrast, \texttt{UCA(Ph1)-18.5.1-RQ2} and \texttt{UCA(Ph1)-18.2.2-RQ1} move by more than five places, indicating that these items can swing from mid-priority to near the bottom and therefore warrant closer investigation.  
\end{itemize}

To deal with such variability, we calculated:

\begin{enumerate}
  \item The \emph{average rank} for every requirement, Equation \eqref{averge rank}, to capture the central tendency across all simulations;  
  \item The \emph{rank variability} (standard deviation), Equation \eqref{standard deviation};  
  \item The \emph{Requirement score} in Equation \eqref{RS}, which combines centrality and variability to provide an objective final rank.  
\end{enumerate}

This approach addresses \textbf{RQ2} (i.e., How can we objectively prioritise all the requirements identified by STPA?): how to incorporate expert judgement without allowing individual perspective to impact the overall analysis. By running the \emph{What if ?} scenarios using MCS and weighting each requirement’s mean position against its spread, the method minimises subjectivity in expert input and limits its influence on the final prioritisation.

Combining the \emph{Requirement score}, based on cost, time, type, and likelihood, with the \emph{UCA-priority score} generates a ranking that is both commercially sensible and safety-focused. However, factors such as implementation cost, expected effort, and regulatory novelty alone are not sufficient to justify a “high-priority” requirement. A requirement exists to mitigate a UCA whose potential losses can be catastrophic; therefore, including the UCA’s own priority score adds a second layer of risk mitigation. By multiplying the Requirement score (commercial effort) by the UCA-priority score (risk), we produce a ranking that involves both business reality and safety urgency.

Figure~\ref{fig: Prioritisation Matrix of UCAs} illustrates a snapshot of the most critical UCAs (labelled P1 and P2), thus, it supplies the risk-related factor in the ranking along with the Requirement score.

For clearer communication, a \emph{requirement-filtering} step was introduced to merge duplicate requirements. The output of the step includes an exported Excel with a list of unique requirements as shown in Table~\ref{tab:reqFiltring}. The same table also illustrates the connection between the unique requirements and their CFs and UCAs. This step enhances readability and streamlines the management of results.

Figure~\ref{fig: Prioritisation Matrix of Requirements} plots the filtered and prioritised set of requirements on two axes, making it easy to visualise requirements with different priorities. 

Requirement \texttt{UCA(Ph0.1)-13.5.2-RQ1}, which falls in the dark-red zone, implies low-cost (\(<30\,\%\)), minor-effort, Type~A, not mitigated by any current regulation, and linked to a highly critical UCA. its position within this area enables stakeholders to readily interpret its priority level with its associated factors.

This presentation style of the results resolved the \textbf{RQ3} (i.e., How can the prioritised requirements be better communicated with the stakeholders?) to facilitate readability of the results.
  

Cells are colour-coded to indicate importance. Consequently, the matrix rescales itself dynamically to any new dataset: the range between “high” and “low” priority expands or contracts with each iteration, making the visual output interactive rather than static. This option addresses \textbf{RQ1}(i.e.,How can the requirements identified from the STPA be effectively implemented for safe system operation?), enabling STPA practitioners to effectively utilise the outputs of STPA.

To address \textbf{RQ3} (i.e., How can the prioritised requirements be better communicated with the stakeholders?), we supply two complementary artefacts that facilitate result communication:

\begin{enumerate}
  \item A dynamically scaled graphical Requirement Matrix, providing direct visualisation of the results; 
  \item An Excel workbook that contains the unique requirements, along with their unique Requirements IDs, which incorporate the related UCA IDs to enhance the traceability. It also includes additional details that cannot be represented in the matrix, enabling quick review and implementation.
\end{enumerate}

Together, these artefacts give stakeholders a balanced view of commercial feasibility and safety urgency, and provide a traceable path from each critical UCA to its mitigating requirement. In \cite{caa_stpa_evtol}, a presentation of the highest priority requirements, those identified as the regulation gap.

The STPA analysis provided a total of 432 requirements. Communicating such a large set to stakeholders, particularly when the goal is to translate them into concrete regulations and policies, poses a challenge. Simply presenting all 432 items, many of which are duplicates, would be time-consuming and incompatible with the strict deadlines of the case study.

To highlight the most relevant findings, we first \emph{rank} the requirements.  
Ranking alone, however, is not sufficient: if several entries represent the same underlying requirement, reviewers must still encounter redundant information.

We therefore introduced a filtering procedure that removes duplicate requirements while retaining their associated causal factors and UCAs.  
The output of this step is illustrated in Table~\ref{tab:reqFiltring}.

Applying this procedure reduced the original set from 432 entries to \textbf{202 unique requirements}, greatly streamlining both communication and subsequent implementation efforts.

This work was conducted as part of an industrial eVTOL project\cite{caa_stpa_evtol}, aiming to address identified research gaps and introduce a novel concept for requirement prioritisation and management. It is not claimed to be definitive or universally applicable, as it was only validated on an eVTOL case study. Further studies on additional systems and operational contexts are needed to demonstrate the method’s broader efficiency and robustness.

\section{Conclusion}
A novel framework to enable the prioritisation of STPA requirements was presented in this paper. It utilizes the outputs from each step of STPA, including the initially identified Losses (from Step 1), the Control Structure (from Step 2), and the UCAs (from Step 3). This allows the prioritisation process to be initiated right from the beginning of STPA. Apart from considering the rankings from safety analysts, the framework also considers inputs from domain experts (i.e., SME), ensuring that a comprehensive set of rankings is collected for the prioritisation. Whilst there can be subjectivity and bias from the individual SME, the Monte-Carlo Simulation was introduced to arrange all the rankings from the SME to ensure that the final ranking is objective. Managing and calculating all the rankings can be time-consuming. The automation toolchain that was developed from the project optimizes the efficiency and accuracy of the prioritisation process. 

Since the requirement prioritisation framework was developed from a project of eVTOL operation, the framework is still aviation-oriented. As part of future work, the aim is to convert the framework to be applied in other domains, including automotive, maritime, and health care etc.

\section{Acknowledgements}
The work presented in this paper was funded by the UK Civil Aviation Authority (GFA 3549). The authors would also like to thank the WMG center of HVM Catapult and WMG, University of Warwick, UK, for providing the necessary infrastructure for conducting this study. WMG hosts one of the seven centers that together comprise the High-Value Manufacturing Catapult in the UK.
\onecolumn
\small
\begin{longtable}{|
    p{0.1\textwidth} 
  | p{0.25\textwidth} 
  | p{0.16\textwidth} 
  | p{0.18\textwidth} 
  | p{0.05\textwidth} 
  |}
  \caption{Requirement Filtering Output}
  \label{tab:reqFiltring}\\
\hline
\textbf{Req ID} & \textbf{UCA Description} & \textbf{Causal Factor(s)} & \textbf{Req Description} & \textbf{Priority} \\
\hline
\endfirsthead

\hline
\textbf{Req ID} & \textbf{UCA Description} & \textbf{Causal Factor(s)} & \textbf{Requirements} & \textbf{Priority} \\
\hline
\endhead

\cellcolor[HTML]{FF5100}UCA(Ph2)-7.5.2-RQ.5
  & Licensed Aerodrome (Silverstone) provides RF/TransponderSetting (squawk) too late when the eVTOL is already approaching its destination.
  & High workload due to simultaneous management of multiple aircraft.
  & \multirow {2}{0.18\textwidth}{Aerodrome control systems shall implement workload management tools during peak traffic conditions.}
  & ReqP3 \\
\cline{1-3}\cline{5-5}
\cellcolor[HTML]{FF5100} UCA(Ph2)-7.5.3-RQ.2
  & Licensed Aerodrome (Silverstone) provides RF/TransponderSetting (squawk) too late when the eVTOL is already experiencing an in-flight emergency.
  & High levels of stress or fatigue prevent operators from effectively addressing and managing emergencies.
  &  
  & ReqP3 \\
\hline

\cellcolor[HTML]{C30000} UCA(Ph0.1)-13.5.2-RQ1
  & Regulator provides eVTOL Approval too late (by x weeks) when flight is already scheduled and the unapproved slot is undetected.
  & The supplementary documents were submitted on time but, due to server/email or postal delays, never reached the Regulator.
  & \multirow {2}{0.18\textwidth}{Aerodrome \,/ Vertiport shall actively seek confirmation of receipt of all supplementary documents sent to the Regulator.}
  & ReqP1 \\
\cline{1-3}\cline{5-5}
\cellcolor[HTML]{FF5100} UCA(Ph0.1)-14.5.1-RQ1
  & Regulator provides TAS (Temp Airspace Structure) to Licensed Aerodrome too late when there is high traffic density on the planned route.
  & The supplementary documents were submitted on time but, due to server/email or postal delays, never reached the Regulator.
  &  
  & ReqP3 \\
\cline{1-3}\cline{5-5}
\cellcolor[HTML]{FE0000} UCA(Ph0.1)-15.5.1-RQ1
  & Regulator provides TAC (Temporary Airfield Creation) too late for an urgent eVTOL flight requiring a temporary landing field, delaying the schedule.
  & The supplementary documents were submitted on time but, due to server/email or postal delays, never reached the Regulator.
  &  
  & ReqP2 \\
\hline

\cellcolor[HTML]{FFFF00} UCA(Ph1)-18.2.1-RQ1
  & NATS (LHR RADAR) provides OnwardClearance incorrectly (wrong altitude/route/heading/speed) when there is a conflict (close proximity to other aircraft, drones, etc.).
  & Controller is unaware of the potential conflict.
  & NATS shall maintain real-time visibility and accurate data on all traffic in the designated airspace to prevent mid-air collisions.
  & ReqP4 \\
\hline

\cellcolor[HTML]{FFFF00} UCA(Ph0.2)-33.7.2-RQ2
  & Temporary Airfield Management stops contracting too soon when the heliport is still required.
  & The Land Approval was correct but misinterpreted by the team.
  & Temporary Airfield Management shall be trained to correctly interpret Land Approval documents.
  & ReqP4 \\
\hline

\cellcolor[HTML]{00FF00} UCA(Ph1)-18.5.1-RQ2
  & NATS (LHR RADAR) provides OnwardClearance too late when the eVTOL is in ‘Hold’ and its energy is nearly depleted.
  & Clearance was issued on time but delayed in the communications channel.
  & There shall be no delays beyond TBD seconds in the comms link between NATS and the eVTOL.
  & ReqP5 \\
\hline

\cellcolor[HTML]{00FF00} UCA(Ph1)-18.2.2-RQ1
  & NATS (LHR RADAR) provides OnwardClearance incorrectly (route not updated) during emergencies or airspace restrictions.
  & Controllers were unaware of emergencies or restrictions along the planned route.
  & NATS must be provided with up-to-date information on emergencies and restricted airspace.
  & ReqP5 \\
\hline

\cellcolor[HTML]{00FF00} UCA(Ph1)-18.2.2-RQ5
  & NATS (LHR RADAR) provides OnwardClearance incorrectly (route not updated) during emergencies or airspace restrictions.
  & The Commander relies solely on NATS and is unaware of route restrictions.
  & The Commander must verify that clearance meets separation standards and is within aircraft capability.
  & ReqP5 \\
\hline

\cellcolor[HTML]{FFFF00} UCA(Ph0.1)-34.1.1-RQ2
  & Licensed Vertiport does not provide T\&C to the Commander when scheduled; the Commander is unaware of them, risking safety-critical losses.
  & T\&C were issued but lost in a lossy channel.
  & \multirow {2}{0.18\textwidth}{The spam/junk email box shall be checked regularly to ensure no critical messages are missed.}
  & ReqP4 \\
\cline{2-3}\cline{5-5}
\cellcolor[HTML]{00FF00} UCA(Ph0.2)-33.1.2-RQ2
  & Temporary Airfield Management does not provide the contract to Silverstone Heliport when required, and it is not detected.
  & The contract was issued but lost in a lossy channel.
  &  
  & ReqP5 \\
\hline

\cellcolor[HTML]{FF5100}\textbf{UCA(Ph0.2)-10.6.1-RQ2}
 & Aircraft-Handling Unit cannot sense the State-of-Charge with update intervals < TBD s.
 & Sensor latency under extreme temperature.
 & AHU shall implement redundant SoC sensors to guarantee update cycles $ < $TBD s.
 & ReqP3 \\ \hline

\cellcolor[HTML]{FFA400}\textbf{UCA(Ph0.1)-17.1.2-RQ1}
 & Regulator does not update the eVTOL performance database when new variants are certified.
 & Processing backlog leads to outdated tables.
 & Regulator shall auto-ingest certification data within 24 h of variant approval.
 & ReqP3 \\ \hline

\cellcolor[HTML]{FFD100}\textbf{UCA(Ph0.1)-49.5.1-RQ4}
 & Regulator provides slot approval too late ($<$ 48 h) for high-priority flights.
 & Queue-prioritisation algorithm does not flag urgent requests.
 & Slot-request API shall include urgency flags so high-priority flights are processed first.
 & ReqP4 \\ \hline

\end{longtable}
\twocolumn


\bibliographystyle{elsarticle-harv} 

\bibliography{main}

\begin{thebibliography}{37}
\expandafter\ifx\csname natexlab\endcsname\relax\def\natexlab#1{#1}\fi
\providecommand{\url}[1]{\texttt{#1}}
\providecommand{\href}[2]{#2}
\providecommand{\path}[1]{#1}
\providecommand{\DOIprefix}{doi:}
\providecommand{\ArXivprefix}{arXiv:}
\providecommand{\URLprefix}{URL: }
\providecommand{\Pubmedprefix}{pmid:}
\providecommand{\doi}[1]{\href{http://dx.doi.org/#1}{\path{#1}}}
\providecommand{\Pubmed}[1]{\href{pmid:#1}{\path{#1}}}
\providecommand{\bibinfo}[2]{#2}
\ifx\xfnm\relax \def\xfnm[#1]{\unskip,\space#1}\fi
\bibitem[{Allison et~al.(2017)Allison, Revell, Sears and Stanton}]{allison2017systems}
\bibinfo{author}{Allison, C.K.}, \bibinfo{author}{Revell, K.M.}, \bibinfo{author}{Sears, R.}, \bibinfo{author}{Stanton, N.A.}, \bibinfo{year}{2017}.
\newblock \bibinfo{title}{Systems theoretic accident model and process (stamp) safety modelling applied to an aircraft rapid decompression event}.
\newblock \bibinfo{journal}{Safety science} \bibinfo{volume}{98}, \bibinfo{pages}{159--166}.
\bibitem[{Badaoui et~al.(2025)Badaoui, Chen, Elizebeth, Khastgir and Jennings}]{elbadaoui2025structured}
\bibinfo{author}{Badaoui, H.E.}, \bibinfo{author}{Chen, S.}, \bibinfo{author}{Elizebeth, M.J.}, \bibinfo{author}{Khastgir, S.}, \bibinfo{author}{Jennings, P.}, \bibinfo{year}{2025}.
\newblock \bibinfo{title}{A structured framework for prioritizing unsafe control actions in stpa: Case study of evtol operations}.
\newblock \bibinfo{note}{Unpublished manuscript}.
\bibitem[{Boyle(1977)}]{Boyle1977}
\bibinfo{author}{Boyle, P.P.}, \bibinfo{year}{1977}.
\newblock \bibinfo{title}{Options: A monte carlo approach}.
\newblock \bibinfo{journal}{Journal of Financial Economics} \bibinfo{volume}{4}, \bibinfo{pages}{323--338}.
\newblock \DOIprefix\doi{10.1016/0304-405X(77)90013-4}. \bibinfo{note}{finance}.
\bibitem[{Cagliero et~al.(2021)Cagliero, Bellini, Marcatto, Novelli, Monteleone and Mazzocchi}]{cumulativevoting}
\bibinfo{author}{Cagliero, R.}, \bibinfo{author}{Bellini, F.}, \bibinfo{author}{Marcatto, F.}, \bibinfo{author}{Novelli, S.}, \bibinfo{author}{Monteleone, A.}, \bibinfo{author}{Mazzocchi, G.}, \bibinfo{year}{2021}.
\newblock \bibinfo{title}{Prioritising cap intervention needs: An improved cumulative voting approach}.
\newblock \bibinfo{journal}{Sustainability} \bibinfo{volume}{13}, \bibinfo{pages}{3997}.
\bibitem[{Chen et~al.(2020)Chen, Khastgir, Babaev and Jennings}]{chen2020identifying}
\bibinfo{author}{Chen, S.}, \bibinfo{author}{Khastgir, S.}, \bibinfo{author}{Babaev, I.}, \bibinfo{author}{Jennings, P.}, \bibinfo{year}{2020}.
\newblock \bibinfo{title}{Identifying accident causes of driver-vehicle interactions using system theoretic process analysis ({STPA})}, in: \bibinfo{booktitle}{2020 IEEE International Conference on Systems, Man, and Cybernetics (SMC)}, \bibinfo{organization}{IEEE}. pp. \bibinfo{pages}{3247--3253}.
\bibitem[{Chen et~al.(2021)Chen, Khastgir and Jennings}]{chen2021analyzing}
\bibinfo{author}{Chen, S.}, \bibinfo{author}{Khastgir, S.}, \bibinfo{author}{Jennings, P.}, \bibinfo{year}{2021}.
\newblock \bibinfo{title}{Analyzing national responses to covid-19 pandemic using stpa}.
\newblock \bibinfo{journal}{Safety Science} \bibinfo{volume}{138}, \bibinfo{pages}{105195}.
\bibitem[{Chen et~al.(2025)Chen, Niu, Mei and Zhang}]{chen2025hierarchical}
\bibinfo{author}{Chen, Z.}, \bibinfo{author}{Niu, C.}, \bibinfo{author}{Mei, M.}, \bibinfo{author}{Zhang, H.}, \bibinfo{year}{2025}.
\newblock \bibinfo{title}{Hierarchical safety analysis and formal verification for safety-critical systems using stamp and event-b}.
\newblock \bibinfo{journal}{Safety Science} \bibinfo{volume}{184}, \bibinfo{pages}{106744}.
\bibitem[{{Civil Aviation Authority}(2025)}]{caa_stpa_evtol}
\bibinfo{author}{{Civil Aviation Authority}}, \bibinfo{year}{2025}.
\newblock \bibinfo{title}{{STPA-based Safety Analysis for eVTOL Operations}}.
\newblock \URLprefix \url{https://www.caa.co.uk/publication/download/25573}.
\bibitem[{El~Badaoui et~al.(2025)El~Badaoui, Chen, Elizebeth, Nakashima, Khastgir and Jennings}]{UCAPrio}
\bibinfo{author}{El~Badaoui, H.}, \bibinfo{author}{Chen, S.}, \bibinfo{author}{Elizebeth, M.J.}, \bibinfo{author}{Nakashima, T.}, \bibinfo{author}{Khastgir, S.}, \bibinfo{author}{Jennings, P.}, \bibinfo{year}{2025}.
\newblock \bibinfo{title}{A structured framework for prioritizing unsafe control actions in stpa: Case study on evtol operations}.
\newblock \bibinfo{journal}{Safety Science} \bibinfo{note}{Preprint submitted August 14, 2025}.
\bibitem[{Fu and Hu(1995)}]{fu1995sensitivity}
\bibinfo{author}{Fu, M.C.}, \bibinfo{author}{Hu, J.Q.}, \bibinfo{year}{1995}.
\newblock \bibinfo{title}{Sensitivity analysis for monte carlo simulation of option pricing}.
\newblock \bibinfo{journal}{Probability in the Engineering and Informational Sciences} \bibinfo{volume}{9}, \bibinfo{pages}{417--446}.
\bibitem[{Goodwin and Wright(2014)}]{GoodwinWright2014}
\bibinfo{author}{Goodwin, P.}, \bibinfo{author}{Wright, G.}, \bibinfo{year}{2014}.
\newblock \bibinfo{title}{Decision Analysis for Management Judgment}.
\newblock \bibinfo{edition}{5th} ed., \bibinfo{publisher}{Wiley}, \bibinfo{address}{Chichester, UK}.
\bibitem[{Harkleroad et~al.(2013)Harkleroad, Vela and Kuchar}]{harkleroad2013review}
\bibinfo{author}{Harkleroad, E.}, \bibinfo{author}{Vela, A.}, \bibinfo{author}{Kuchar, J.}, \bibinfo{year}{2013}.
\newblock \bibinfo{title}{Review of systems-theoretic process analysis (stpa) method and results to support nextgen concept assessment and validation}.
\newblock \bibinfo{journal}{Proj. Report: ATC-427 MIT} .
\bibitem[{{International Organization for Standardization}(2009)}]{ISO31010}
\bibinfo{author}{{International Organization for Standardization}}, \bibinfo{year}{2009}.
\newblock \bibinfo{title}{{ISO/IEC 31010:2009} Risk management—Risk assessment techniques}.
\newblock \bibinfo{type}{International Standard} \bibinfo{number}{ISO/IEC 31010:2009}. International Organization for Standardization. \bibinfo{address}{Geneva, Switzerland}.
\newblock \bibinfo{note}{Accessed: 23 May 2025}.
\bibitem[{James~Elizebeth et~al.(2023)James~Elizebeth, Khastgir, Babaev, Chen and Jennings}]{comparison}
\bibinfo{author}{James~Elizebeth, M.}, \bibinfo{author}{Khastgir, S.}, \bibinfo{author}{Babaev, I.}, \bibinfo{author}{Chen, S.}, \bibinfo{author}{Jennings, P.}, \bibinfo{year}{2023}.
\newblock \bibinfo{title}{Comparison of fta and stpa approaches: a brake-by-wire case study}.
\newblock \bibinfo{journal}{Siddartha and Babaev, Islam and Chen, Shufeng and Jennings, Paul, Comparison of FTA and Stpa Approaches: A Brake-by-Wire Case Study} .
\bibitem[{Khastgir et~al.(2021)Khastgir, Brewerton, Thomas and Jennings}]{khastgir2021systems}
\bibinfo{author}{Khastgir, S.}, \bibinfo{author}{Brewerton, S.}, \bibinfo{author}{Thomas, J.}, \bibinfo{author}{Jennings, P.}, \bibinfo{year}{2021}.
\newblock \bibinfo{title}{Systems approach to creating test scenarios for automated driving systems}.
\newblock \bibinfo{journal}{Reliability engineering \& system safety} \bibinfo{volume}{215}, \bibinfo{pages}{107610}.
\bibitem[{Khastgir et~al.(2017)Khastgir, Sivencrona, Dhadyalla, Billing, Birrell and Jennings}]{khastgir2017introducing}
\bibinfo{author}{Khastgir, S.}, \bibinfo{author}{Sivencrona, H.}, \bibinfo{author}{Dhadyalla, G.}, \bibinfo{author}{Billing, P.}, \bibinfo{author}{Birrell, S.}, \bibinfo{author}{Jennings, P.}, \bibinfo{year}{2017}.
\newblock \bibinfo{title}{Introducing asil inspired dynamic tactical safety decision framework for automated vehicles}, in: \bibinfo{booktitle}{2017 IEEE 20th International Conference on Intelligent Transportation Systems (ITSC)}, \bibinfo{organization}{IEEE}. pp. \bibinfo{pages}{1--6}.
\bibitem[{Kravchenko et~al.(2022)Kravchenko, Bogdanova and Shevgunov}]{moscow}
\bibinfo{author}{Kravchenko, T.}, \bibinfo{author}{Bogdanova, T.}, \bibinfo{author}{Shevgunov, T.}, \bibinfo{year}{2022}.
\newblock \bibinfo{title}{Ranking requirements using moscow methodology in practice}, in: \bibinfo{booktitle}{Computer Science On-line Conference}, \bibinfo{organization}{Springer}. pp. \bibinfo{pages}{188--199}.
\bibitem[{Kriswardhana et~al.(2025)Kriswardhana, Toaza, Esztergar-Kiss and Duleba}]{ahp}
\bibinfo{author}{Kriswardhana, W.}, \bibinfo{author}{Toaza, B.}, \bibinfo{author}{Esztergar-Kiss, D.}, \bibinfo{author}{Duleba, S.}, \bibinfo{year}{2025}.
\newblock \bibinfo{title}{Analytic hierarchy process in transportation decision-making: A two-staged review on the themes and trends of two decades}.
\newblock \bibinfo{journal}{Expert Systems with Applications} \bibinfo{volume}{261}, \bibinfo{pages}{125491}.
\bibitem[{Kıvanç et~al.(2025)Kıvanç, Tuzkaya and Vayvay}]{Kivanc2025SMSRiskAviation}
\bibinfo{author}{Kıvanç, E.}, \bibinfo{author}{Tuzkaya, G.}, \bibinfo{author}{Vayvay, O.}, \bibinfo{year}{2025}.
\newblock \bibinfo{title}{Safety management system and risk-based approach in aviation maintenance: A systematic literature review}.
\newblock \bibinfo{journal}{Safety Science} \bibinfo{volume}{184}, \bibinfo{pages}{106755}.
\newblock \DOIprefix\doi{10.1016/j.ssci.2024.106755}.
\bibitem[{Leveson(2016)}]{leveson2016engineering}
\bibinfo{author}{Leveson, N.G.}, \bibinfo{year}{2016}.
\newblock \bibinfo{title}{Engineering a safer world: Systems thinking applied to safety}.
\newblock \bibinfo{publisher}{The MIT Press}.
\bibitem[{Nakashima et~al.(2025)Nakashima, Kureta and Khastgir}]{nakashima2025addressing}
\bibinfo{author}{Nakashima, T.}, \bibinfo{author}{Kureta, R.}, \bibinfo{author}{Khastgir, S.}, \bibinfo{year}{2025}.
\newblock \bibinfo{title}{Addressing systemic risks in autonomous maritime navigation: A structured stpa and odd-based methodology}.
\newblock \bibinfo{journal}{Reliability Engineering and System Safety} \bibinfo{volume}{261}.
\bibitem[{Olaru et~al.(2014)Olaru, {\c{S}}andru and Pirnea}]{olaru2014monte}
\bibinfo{author}{Olaru, M.}, \bibinfo{author}{{\c{S}}andru, M.}, \bibinfo{author}{Pirnea, I.C.}, \bibinfo{year}{2014}.
\newblock \bibinfo{title}{Monte carlo method application for environmental risks impact assessment in investment projects}.
\newblock \bibinfo{journal}{Procedia-Social and Behavioral Sciences} \bibinfo{volume}{109}, \bibinfo{pages}{940--943}.
\bibitem[{Pidd(2009)}]{Pidd2009}
\bibinfo{author}{Pidd, M.}, \bibinfo{year}{2009}.
\newblock \bibinfo{title}{Tools for Thinking: Modelling in Management Science}.
\newblock \bibinfo{edition}{2nd} ed., \bibinfo{publisher}{Wiley}, \bibinfo{address}{Chichester, UK}.
\bibitem[{Qi et~al.(2025)Qi, Zhao, Khastgir and Huang}]{qi2025safety}
\bibinfo{author}{Qi, Y.}, \bibinfo{author}{Zhao, X.}, \bibinfo{author}{Khastgir, S.}, \bibinfo{author}{Huang, X.}, \bibinfo{year}{2025}.
\newblock \bibinfo{title}{safety analysis in the era of large language models: a case study of stpa using chatgpt}.
\newblock \bibinfo{journal}{Machine Learning with Applications} , \bibinfo{pages}{100622}.
\bibitem[{Riccardi et~al.(2025)Riccardi, Compare, Mascherona and Zio}]{riccardi2025structural}
\bibinfo{author}{Riccardi, L.}, \bibinfo{author}{Compare, M.}, \bibinfo{author}{Mascherona, R.}, \bibinfo{author}{Zio, E.}, \bibinfo{year}{2025}.
\newblock \bibinfo{title}{Structural causal modeling and stpa for the risk analysis of a rail system powered by h2 fuel}.
\newblock \bibinfo{journal}{Reliability Engineering \& System Safety} \bibinfo{volume}{256}, \bibinfo{pages}{110758}.
\bibitem[{Sadeghi and Goerlandt(2023)}]{sadeghi2023proposed}
\bibinfo{author}{Sadeghi, R.}, \bibinfo{author}{Goerlandt, F.}, \bibinfo{year}{2023}.
\newblock \bibinfo{title}{A proposed validation framework for the system theoretic process analysis (stpa) technique}.
\newblock \bibinfo{journal}{Safety science} \bibinfo{volume}{162}, \bibinfo{pages}{106080}.
\bibitem[{Saltelli et~al.(2008)}]{saltelli2008}
\bibinfo{author}{Saltelli, A.}, et~al., \bibinfo{year}{2008}.
\newblock \bibinfo{title}{Global Sensitivity Analysis: The Primer}.
\newblock \bibinfo{publisher}{John Wiley \& Sons}.
\bibitem[{Schmid and Stanton(2018)}]{schmid2018laser}
\bibinfo{author}{Schmid, D.}, \bibinfo{author}{Stanton, N.A.}, \bibinfo{year}{2018}.
\newblock \bibinfo{title}{How are laser attacks encountered in commercial aviation? a hazard analysis based on systems theory}.
\newblock \bibinfo{journal}{Safety science} \bibinfo{volume}{110}, \bibinfo{pages}{178--191}.
\bibitem[{Slevitch(2025)}]{kano}
\bibinfo{author}{Slevitch, L.}, \bibinfo{year}{2025}.
\newblock \bibinfo{title}{Kano model categorization methods: Typology and systematic critical overview for hospitality and tourism academics and practitioners}.
\newblock \bibinfo{journal}{Journal of Hospitality \& Tourism Research} \bibinfo{volume}{49}, \bibinfo{pages}{449--479}.
\bibitem[{Tasneem et~al.(2025)Tasneem, Zulzalil and Hassan}]{tasneem2025enhancing}
\bibinfo{author}{Tasneem, N.}, \bibinfo{author}{Zulzalil, H.B.}, \bibinfo{author}{Hassan, S.}, \bibinfo{year}{2025}.
\newblock \bibinfo{title}{Enhancing agile software development: A systematic literature review of requirement prioritization and reprioritization techniques}.
\newblock \bibinfo{journal}{IEEE Access} .
\bibitem[{Thomas et~al.(2024)Thomas, Van~Houdt et~al.}]{thomas2024evaluation}
\bibinfo{author}{Thomas, J.P.}, \bibinfo{author}{Van~Houdt, J.G.}, et~al., \bibinfo{year}{2024}.
\newblock \bibinfo{title}{Evaluation of System-Theoretic Process Analysis (STPA) for Improving Aviation Safety}.
\newblock \bibinfo{type}{Technical Report}. United States. Department of Transportation. Federal Aviation Administration~….
\bibitem[{Tonk and Boussif(2024)}]{tonk2024application}
\bibinfo{author}{Tonk, A.}, \bibinfo{author}{Boussif, A.}, \bibinfo{year}{2024}.
\newblock \bibinfo{title}{Application of systems theoretic accident model and processes in railway systems: A review}.
\newblock \bibinfo{journal}{IEEE Access} .
\bibitem[{Ventikos et~al.(2020)Ventikos, Chmurski and Louzis}]{ventikos2020systems}
\bibinfo{author}{Ventikos, N.P.}, \bibinfo{author}{Chmurski, A.}, \bibinfo{author}{Louzis, K.}, \bibinfo{year}{2020}.
\newblock \bibinfo{title}{A systems-based application for autonomous vessels safety: Hazard identification as a function of increasing autonomy levels}.
\newblock \bibinfo{journal}{Safety science} \bibinfo{volume}{131}, \bibinfo{pages}{104919}.
\bibitem[{Wong et~al.(2020)Wong, Huynh, Mak, Leveson and Singer}]{wong2020stamping}
\bibinfo{author}{Wong, L.}, \bibinfo{author}{Huynh, E.}, \bibinfo{author}{Mak, R.}, \bibinfo{author}{Leveson, N.}, \bibinfo{author}{Singer, L.}, \bibinfo{year}{2020}.
\newblock \bibinfo{title}{Stamping out mri simulation hazards with a system-theoretic accident model and processes approach to proactive hazard assessment}.
\newblock \bibinfo{journal}{International Journal of Radiation Oncology, Biology, Physics} \bibinfo{volume}{108}, \bibinfo{pages}{e204--e205}.
\bibitem[{Yamaguchi and Thomas(2019)}]{yamaguchi2019system}
\bibinfo{author}{Yamaguchi, S.}, \bibinfo{author}{Thomas, J.}, \bibinfo{year}{2019}.
\newblock \bibinfo{title}{A system safety approach for tomographic treatment}.
\newblock \bibinfo{journal}{Safety Science} \bibinfo{volume}{118}, \bibinfo{pages}{772--782}.
\bibitem[{Yang et~al.(2020)Yang, Utne, Sand{\o}y, Ramos and Rokseth}]{yang2020systems}
\bibinfo{author}{Yang, X.}, \bibinfo{author}{Utne, I.B.}, \bibinfo{author}{Sand{\o}y, S.S.}, \bibinfo{author}{Ramos, M.A.}, \bibinfo{author}{Rokseth, B.}, \bibinfo{year}{2020}.
\newblock \bibinfo{title}{A systems-theoretic approach to hazard identification of marine systems with dynamic autonomy}.
\newblock \bibinfo{journal}{Ocean Engineering} \bibinfo{volume}{217}, \bibinfo{pages}{107930}.
\bibitem[{Zhou et~al.(2019)Zhou, Zi, Chen and An}]{zhou2019hazard}
\bibinfo{author}{Zhou, Z.}, \bibinfo{author}{Zi, Y.}, \bibinfo{author}{Chen, J.}, \bibinfo{author}{An, T.}, \bibinfo{year}{2019}.
\newblock \bibinfo{title}{Hazard analysis for escalator emergency braking system via system safety analysis method based on stamp}.
\newblock \bibinfo{journal}{Applied Sciences} \bibinfo{volume}{9}, \bibinfo{pages}{4530}.

\end{thebibliography}






\end{document}